# Eigenmodes of Néel skyrmions in ultrathin magnetic films

Vanessa Li Zhang,[1] Chen Guang Hou,[1] Kai Di,[1] Hock Siah Lim,[1] Ser Choon Ng,[1] Shawn David Pollard,[2] Hyunsoo Yang,[2] and Meng Hau Kuok[1,a]

[1]*Department of Physics, National University of Singapore, Singapore 117551*

[2]*Department of Electrical and Computer Engineering, National University of Singapore, Singapore 117576*

**The static and dynamic states of Néel skyrmions in ultrathin ferromagnetic films with interfacial Dzyaloshinskii-Moriya interaction (DMI) have been micromagnetically simulated as functions of the interfacial DMI strength and applied static magnetic field. Findings reveal that while the breathing, counterclockwise (CCW) and clockwise (CW) rotational eigenmodes exist in the skyrmion lattice (SkL) phase, only the first two modes are present in the isolated skyrmion (ISk) phase. Additionally, the eigenfrequency of the CCW mode is insensitive to the magnetic-field driven SkL–ISk phase transition, and the inter-skyrmion interaction is largely responsible for exciting the CW mode in the SkL phase. The findings provide physical insight into the dynamics of the phase transition and would be of use to potential skyrmion-based microwave applications.**

[a] phykmh@nus.edu.sg

**INTRODUCTION**

Because of its rich fundamental physics and enormous technological application potential, the magnetic skyrmion is currently a very hot research topic [1-7]. First discovered in 2009 [3], this remarkable magnetic entity is a chiral spin texture with extraordinary attributes such as its nanometric size, topological stability and ability to propagate under ultralow current densities [2,5]. Envisaged as building blocks for novel spintronic devices, skyrmions are also anticipated to be information carrying bits in next-generation low-energy, ultra-compact data storage and processing devices [6,7].

Skyrmions are topologically protected spin textures which often originate from chiral interactions known as Dzyaloshinskii-Moriya interactions (DMI) [8,9]. Investigations into their static spin structures reveal that the bulk DMI stabilizes vortex-type (or Bloch-type) skyrmions in B20 crystallographic structures [3,10], while the interfacial DMI leads to the formation of hedgehog-type (or Néel-type) skyrmions in ultrathin ferromagnetic films deposited on a strong spin-orbit metal [5-7,10,11,12]. Extensive studies have been done on the collective dynamics of skyrmions under external fields. Among them are investigations into the current-induced motion of skyrmions and the associated skyrmion Hall effect [7,13,14], the collective rotation of the skyrmion lattice [15], and the pinning of skyrmions [16]. In contrast, the dynamics of the internal structure of the skyrmion has not been intensively explored [17-20]. This is particularly so for the Néel skyrmion whose internal dynamics is much less researched than that of its Bloch counterpart. Three ac magnetic-field-induced fundamental collective eigenmodes of the Bloch skyrmion lattice were theoretically predicted by Mochizuki [17]. Based on their real space configurations, they are labelled the breathing mode, the clockwise (CW) and the counterclockwise (CCW) rotational modes. The existence of the breathing and CCW skyrmion eigenmodes has been experimentally verified by microwave transmittance spectroscopic studies of bulk $Cu_2OSeO_3$ [18,21]. Very recently, the experimental observation of the three fundamental collective spin excitations of the Néel skyrmion lattice in $GaV_4S_8$ single crystals was reported [22]. It is worth noting that simulations performed by Liu *et al*. [23] have unveiled a rich variety of internal modes of the Bloch skyrmion excited by an ac electric field in the multiferroic $Cu_2OSeO_3$ crystal.

As the eigenmodes of skyrmions provide information on their stability and rigidity, an elucidation of the eigenmodes is important for understanding the behavior of these magnetic quasiparticles. Knowledge of the size of skyrmions and the skyrmion lattice constant is crucial to skyrmion-based device applications, as these parameters have to be tuned for maximum performance. Additionally, information on the magnetic field dependence of the eigenmodes provides guidance to the detection and characterization of skyrmions in experiments.

Recent attention has focused on ultrathin ferromagnet/heavy-metal films, as they exhibit strong interfacial DMI which support skyrmions [24-29]. Here, we report on the theoretical study of the low-frequency eigen-excitations of the Néel skyrmion in ultrathin ferromagnetic films as functions of the DMI strength, and an applied static magnetic field. The dependence on these quantities of individual skyrmion sizes of isolated skyrmions and skyrmion lattices, as well as the lattice constants of the latter were also investigated. The dynamics of the skyrmion lattice was also simulated through the skyrmion-lattice to isolated-skyrmion phase transition.

**SIMULATIONS**

The modeled system is a 0.8nm-thick CoFeB film with an interfacial DMI. CoFeB is a commonly used ferromagnetic component in ultrathin heavy-metal/ferromagnet multilayer films [26,30], as such structures exhibit large DMIs which is the usual requisite for stabilizing skyrmions. Micromagnetic simulations using the MuMax3 program [31], were performed to investigate the static and dynamic properties of Néel skyrmions in the film by solving the Landau-Lifshitz-Gilbert equation

$$\frac{d\mathbf{m}}{dt} = -\gamma \mathbf{m} \times \mathbf{H}_{eff} + \alpha \mathbf{m} \times \frac{d\mathbf{m}}{dt}, \quad (1)$$

where $\gamma = 190$ GHz/T is the gyromagnetic ratio, $\alpha = 0.01$ the Gilbert damping parameter, and $\mathbf{m}$ the unit vector of the magnetization. The effective field $\mathbf{H}_{eff}$ is given by

$$\mathbf{H}_{eff} = H_0 \hat{e}_z + \frac{2A}{\mu_0 M_S} \nabla^2 \mathbf{m} + \mathbf{H}_{demag} + \frac{2K_U}{\mu_0 M_S} m_z \hat{e}_z + \frac{2D}{\mu_0 M_S} \left[ \frac{\partial m_z}{\partial x} \hat{e}_x + \frac{\partial m_z}{\partial y} \hat{e}_y - \left( \frac{\partial m_x}{\partial x} + \frac{\partial m_y}{\partial y} \right) \hat{e}_z \right],$$

(2)

where $H_0$ is the static magnetic field applied normal to the film plane, $\mu_0$ the vacuum permeability, $A = 16 \times 10^{-12}$ J/m the exchange stiffness constant, $M_S = 1.6 \times 10^6$ A/m the saturation magnetization, $H_{demag}$ the demagnetizing field, $K_U = 2 \times 10^6$ J/m$^3$ the uniaxial anisotropy constant, and $D$ the DMI constant $\hat{e}_x$, $\hat{e}_y$, and $\hat{e}_z$ are the unit vectors of the Cartesian coordinate system, in which the z-axis is normal to the film plane. The above magnetic parameters are experimental values [26,29], except for the DMI constant, which was varied from $D = 1$ to 6 mJ/m$^2$ in the simulations. It is to be noted that Yang *et al*. have established that a value of $D$ as high as 7 mJ/m$^2$ is theoretically feasible [32].

The $D$-$H_0$ phase diagram was evaluated as follows. The ground states were obtained by slowly cooling the modeled spin system, with a random magnetization configuration, from a high temperature to 0 K, by including a white-noise thermal field that is proportional to the square root of temperature, in the effective field $H_{eff}$ [31]. The computational cell is a cuboid, with periodic boundary conditions imposed along the x- and y-axes (Fig. 1(b) shows the x-y plane of the cell), and magnetization assumed to be uniform along its thickness of 0.8 nm. Four magnetic phases, namely the domain wall (DW), hexagonal skyrmion lattice (SkL), isolated skyrmion (ISk), and ferromagnetic (FM) phases were found. The dynamics of the eigenmodes of skyrmions in the hexagonal SkL and ISk phases were next investigated. For the former, the computational cell used encompasses two skyrmions, and is of dimensions $\sqrt{3}a \times a \times 0.8\,\text{nm}$. The lattice constant $a$ of the ground-state skyrmion crystal was determined by minimizing the total surface magnetic energy density, which was obtained by relaxing a preset skyrmion-like initial magnetization distribution. However, there are cases where a minimal energy does not exist for finite values of $a$, corresponding to the preclusion of a stable lattice, i.e. to the presence of the ISk phase. A computational cell, with a planar square surface, containing a single skyrmion at its center, was used for modeling the ISk phase, with the imposition of the same periodic boundary conditions. A finite-difference cell size of 1 nm × 1 nm × 0.8 nm was used in simulations.

The frequencies and phase distributions of dynamic modes were then evaluated by obtaining the Fourier spectrum of the magnetization distribution driven by an excitation magnetic field pulse applied either parallel or perpendicular to the $z$-axis. This field, which is a sinc function of frequency 100 GHz, can excite eigenmodes with frequencies up to 100 GHz. To exclusively excite a specific mode of frequency $\omega$, a sinusoidal magnetic field $\mathbf{H}_\omega$ with the same frequency $\omega$ was applied, with $\mathbf{H}_\omega$ directed out-of-plane and in-plane to excite the breathing and rotational modes, respectively. Its amplitude, which was typically 0.1% that of the static field $H_0$, was kept sufficiently low to preclude melting of the skyrmion lattice.

## RESULTS AND DISCUSSIONS

The simulated phase diagram, presented in Fig. 1(a), reveals that the 0 K magnetic ground state can assume one of the DW, SkL, ISk, and FM phases. Representative magnetization configurations of the DW, SkL (lattice constant $a$ = 49 nm), and ISk phases at $H_0$ = 100, 325, and 500 mT/$\mu_0$ respectively, are depicted in Fig. 1(b) for the $D$ = 6 mJ/m$^2$ film.

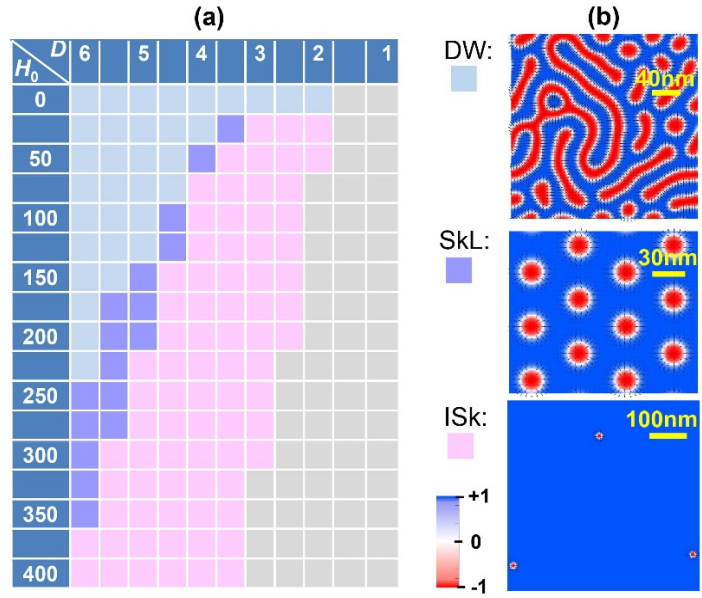

**FIG. 1.** (a) Phase diagram of the ultrathin ferromagnetic films in the plane of applied magnetic field $H_0$ (mT/$\mu_0$) and DMI constant $D$ (mJ/m$^2$). (b) Representative magnetization configurations, for the $D$ = 6 mJ/m$^2$ film under various $H_0$ fields, of the domain wall (DW), skyrmion lattice (SkL) and isolated skyrmion (ISk) phases. The arrows represent the in-plane magnetization components $\mathbf{m}_x + \mathbf{m}_y$, while the out-of-plane components $m_z$ are color coded. The ferromagnetic phase is denoted by grey squares.

The dependences of the properties of the skyrmion lattice on applied magnetic field are displayed in Fig. 2. Figure 2(a) reveals that, for the same field, the stronger the DMI, the shorter is the skyrmion lattice constant $a$; while for a given $D$, the lattice expands with increasing field. For a large $D$ of 6 mJ/m$^2$, $a$ can be as small as 43 nm, while for a moderate $D$ of 3.5 mJ/m$^2$, $a$ can be as large as 110 nm. Systems with $D \geq 3.5$ mJ/m$^2$ can host skyrmion crystals over respective ranges of $H_0$, within which, while their hexagonal Bravais symmetry is maintained, their lattice constants $a$ do vary. Generally, the variation of $a$ with increasing $H_0$ is gradual, but it changes abruptly when $H_0$ approaches the SkL–ISk phase transition value, which for example is about 200 mT/$\mu_0$ for the $D = 5$ mJ/m$^2$ sample.

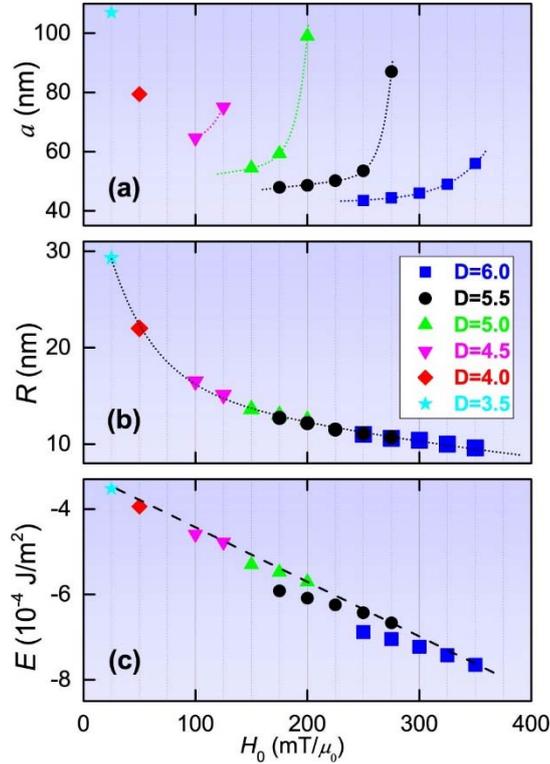

**FIG. 2.** Magnetic field dependences of (a) lattice constant, (b) skyrmion radius and (c) total surface magnetic energy density in the skyrmion lattice phase. The dotted curves are a guide to the eye. The dashed line in (c) represents the surface magnetic energy density of the system in the ferromagnetic phase.

Typically, the lattice constant *a* is a few times larger than the skyrmion radius, which is of the order of tens of nanometers. The skyrmion radius *R* is taken to be the radius of the circle corresponding to $m_z = 0$ [33,34]. The magnetic field dependence of the radius, for different *D* values, is represented by the smooth monotonic curve of Fig. 2(b). The decrease in radius with increasing field is consistent with the findings of Romming *et al*. [34] in their spin-polarized scanning tunneling microscopic study of isolated interface-induced skyrmions. As the applied field $H_0$ favors uniform alignment of spins, the stronger it is, the smaller will be the skyrmion, until eventually for sufficiently large fields, the film adopts the FM phase. The dashed line in Fig. 2(c) represents the surface magnetic energy densities of the systems in the FM phase as a function of $H_0$. For any given $H_0$, the energy density of the SkL phase is lower than that of the corresponding FM phase, thus indicating that the skyrmion lattice is the ground state.

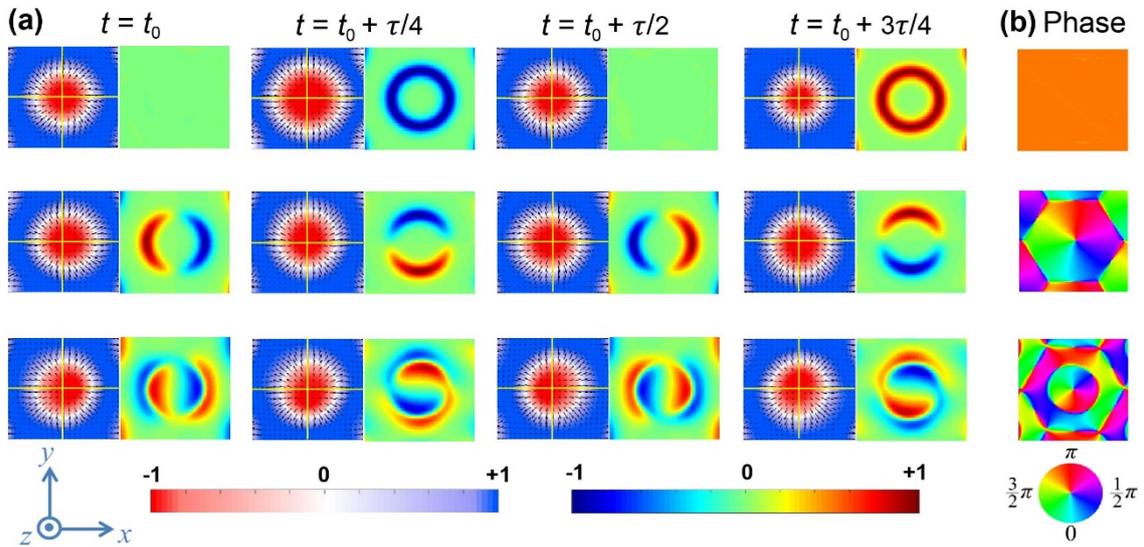

**FIG. 3.** Eigenmodes of a skyrmion in the SkL phase for the $D = 4$ mJ/m$^2$ film under a magnetic field $\boldsymbol{H} = \boldsymbol{H}_0 + \boldsymbol{H}_\omega$, where $H_0 = 50$ mT/$\mu_0$, and $H_\omega = 0.001 H_0 \sin\omega t$. The mode frequencies $\omega$ are 4.4, 8.8 and 22.8 GHz for the breathing (top row), CCW (middle row)

and CW (bottom row) modes, respectively. (a) Snapshots, over one period $\tau$, of the mode profiles (left subpanels), and the corresponding dynamic magnetization profiles [$m_z(t) - m_z(t < 0)$] (right subpanels), where $m_z(t < 0)$ is the value of $m_z$ prior to the application of $H_\omega$. The arrows represent the in-plane spin components $\mathbf{m}_x + \mathbf{m}_y$, while the out-of-plane spin components $m_z$ are color coded (left color bar). The dynamic magnetization profiles are color coded according to the right color bar. (b) Phase angles of $m_z$ for the respective eigenmodes.

Presented in Fig. 3 are the three eigenmodes computed for $D = 4$ mJ/m$^2$ and $H_0 = 50$ mT/$\mu_0$. The temporal evolutions, over one period $\tau$, of the magnetization profiles of a constituent skyrmion in a SkL are displayed in Fig. 3(a). Also shown alongside the profiles, are the respective profiles of the dynamic magnetization [$m_z(t) - m_z(t < 0)$], where $m_z(t < 0)$ is the value of $m_z$ prior to the application of the excitation field $H_\omega$. The characteristic motions of the three modes are clearly manifested in the respective temporal evolutions of their dynamic magnetization profiles. Also, the rotation senses (or lack thereof) of all three modes are clearly depicted by the snapshots, over one cycle, of their respective profiles presented in the figure. Figure 3(b) illustrates the constant phase angle of the breathing mode, and the rotation in the clockwise (counterclockwise) sense of the phase angle of the CW (CCW) mode. Mruczkiewicz *et al.* recently formulated a classification of the fundamental breathing, CCW and CW eigenexcitations based on their azimuthal and radial mode indices ($m = 0$, $n = 0$), ($m = +1$, $n = 0$) and ($m = -1$, $n = 1$), respectively [35]. Noteworthily, the numbers of radial nodes $n$, and the rotation senses of the simulated modes, namely $m = 0$ (breathing; absence of rotation), +1 (CCW) and −1 (CW), are consistent with the classification. Figure 3 shows that the dynamic magnetization and phase profiles of the CW mode exhibit a node along the radial direction of the skyrmion, while the profiles of the other two excitations are devoid of such nodes.

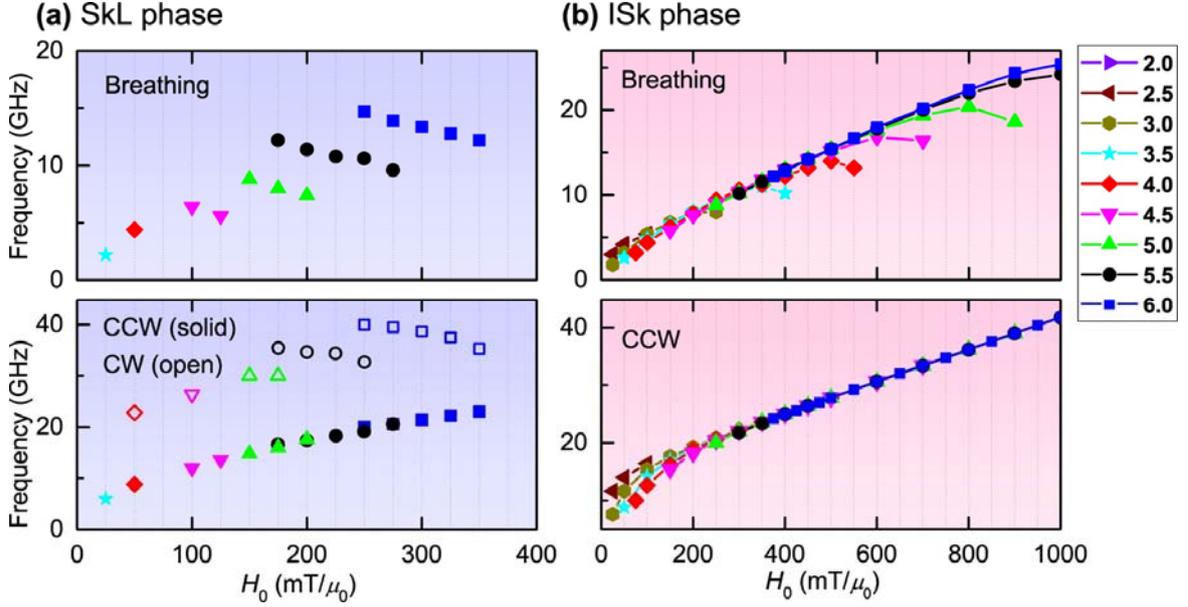

**FIG. 4.** Magnetic field dependence of the resonant frequencies of the eigenmodes of (a) a skyrmion in the SkL phase and (b) an isolated skyrmion for various values of the DMI constant. The curves are a guide to the eye.

The variations of the frequencies of the fundamental eigenexcitations in the SkL and ISk phases, with applied magnetic field are summarized in Fig. 4. A common feature is, for a given film (i.e. fixed $D$) in each phase, the general trend of the field dependence of the frequency is the same for each of the modes. For instance, Fig. 4(a) shows that, in the SkL phase, the frequencies of the breathing and CW modes exhibit a falling field dependence, while that of the CCW mode rises with increasing field. In the case of the ISk, the field dependence of breathing mode frequency, for the various $D$ values, is represented by an upward curve that is generally smoothly varying, except in the regions where mode softening occurs, corresponding to the onset of the transition to the ferromagnetic phase, and eventually resulting in the disappearance of the skyrmion spin textures, as shown in Fig. 4(b). Similarly, the field dependence of the CCW mode frequency in the ISk phase appears as a smoothly varying upward curve in the figure. Additionally, it is observed from Fig. 4(a) that the frequencies of the breathing and CW modes of the skyrmion crystal increase with increasing $D$ for fixed values of the field $H_0$.

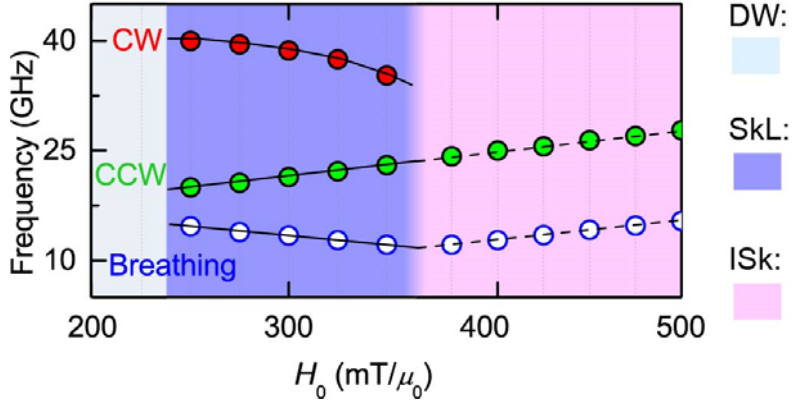

**FIG. 5.** Magnetic phase transition in the modeled ferromagnetic film with $D = 6$ mJ/m$^2$. The skyrmion-lattice (SkL) to isolated-skyrmion (ISk) phase transition occurs at approximately $H_0 = 360$ mT/$\mu_0$. The curves are a guide to the eye.

It is of note that the computed field dependences of the three collective mode frequencies of the SkL for the $D = 6$ mJ/m$^2$ film, presented in Fig. 5, are similar to those experimentally observed for Néel skyrmion lattices hosted in bulk single crystals of GaV$_4$S$_8$ [22]. The simulations also confirm the experimental energy hierarchy of the fundamental modes [36]. Additionally, the figure reveals that the CCW mode is unaffected by the SkL–ISk phase transition, as its frequency exhibits the same linear magnetic field dependence in both phases. The CW and breathing modes, on the other hand, are sensitive to this transition. Specifically, the ISk does not execute the CW mode, while its breathing mode frequency rises linearly with increasing field, in contrast to the falling dependence in the SkL phase. As the CCW and CW modes have opposite rotation senses, relative to the applied field, their dependences on the inter-skyrmion interaction are deduced to be different as follows. The observation that the former is unaffected by the phase transition implies that the interaction has an insignificant influence on its excitation, as the interaction is more pronounced in the SkL than in the ISk phase. In contrast, the CW mode is present only in the SkL phase, suggesting that the inter-skyrmion interaction is mainly responsible for exciting the CW mode in this phase. In this connection, it is noteworthy that in the SkL region of Fig. 4(a), there are for the same $D$ value, correspondingly fewer CW data points than CCW ones. This is because in the vicinity of the phase transition, the inter-skyrmion interaction weakens drastically with sharply increasing skyrmion-skyrmion separation [see Fig. 2(a)], eventually resulting in the ceasing of the CW mode excitation beyond the transition point. Also, it should be noted that the inter-skyrmion separation for ISk is very much larger than that in the SkL phase. This can be seen from Fig. 2(a), where e.g. for $D$ = 5.5 mJ/m$^2$, starting from $H_0 \approx 175$ mT/$\mu_0$, the lattice constant $a$ increases gradually with field until at $H_0 \approx 290$ mT/$\mu_0$, $a$ increases very sharply, signaling the transition to the ISk phase. Figure 1(b) shows that the skyrmion-skyrmion spacing in the ISk phase is several times larger than that in the SkL one. Other possible reasons for the absence of the CW mode in the ISk phase include a change in the internal structure of the skyrmion, arising from the increase in the magnetic field, namely that the skyrmion internal structure in the SkL phase is less symmetrical than that in the ISk phase. Finally, it should be pointed out that because of the instability of the modes in the vicinity of the transition, data points within this region were not obtained, and thus the demarcation of the transition, occurring at approximately $H_0 = 360$ mT/$\mu_0$, is not a sharp one as illustrated in Fig. 5.

**CONCLUSIONS**

In summary, micromagnetic simulations based on MuMax3 have been performed to ascertain the dependence of the static and dynamic Néel skyrmion states, in ultrathin ferromagnetic films, on interfacial DMI strength and applied magnetic field. The simulations corroborate the recently-formulated classification of the three fundamental Néel skyrmion eigenmodes [35], and also confirm the experimental energy hierarchy of these modes in $GaV_4S_8$ [36]. Of particular interest is the observation that the three modes exhibit drastically different magnetic-field dependences across the skyrmion-lattice to isolated-skyrmion phase transition. It is found that the inter-skyrmion interaction is largely responsible for the existence of the CW mode in the lattice phase. The findings provide physical insight into the dynamics of the phase transition, as well as guidance to the experimental identification of the skyrmion modes in ultrathin ferromagnetic films with the DMI. Additionally, as the skyrmions possess eigenfrequencies in the gigahertz range, they show promise for microwave applications such as skyrmion-based tunable resonators.

**ACKNOWLEDGMENT**

Support from the Ministry of Education, Singapore under Academic Research Grant No. R144-000-371-114 is gratefully acknowledged. CGH and DK have contributed equally to this work.

1. U. K. Roessler, A. N. Bogdanov, and C. Pfleiderer, Nature **442**, 797 (2006).

2. S. Heinze, K. von Bergmann, M. Menzel, J. Brede, A. Kubetzka, R. Wiesendanger, G. Bihlmayer, and S. Blugel, Nat. Phys. **7**, 713 (2011).

3. S. Muhlbauer, B. Binz, F. Jonietz, C. Pfleiderer, A. Rosch, A. Neubauer, R. Georgii, and P. Boni, Science **323**, 915 (2009).

4. I. Kezsmarki, S. Bordacs, P. Milde, E. Neuber, L. M. Eng, J. S. White, H. M. Ronnow, C. D. Dewhurst, M. Mochizuki, K. Yanai, H. Nakamura, D. Ehlers, V. Tsurkan and A. Loidl, Nat. Mater. **14**, 1116 (2015).

5. J. Sampaio, V. Cros, S. Rohart, A. Thiaville, and A. Fert, Nat. Nanotech. **8**, 839 (2013).

6. X. Z. Yu, Y. Onose, N. Kanazawa, J. H. Park, J. H. Han, Y. Matsui, N. Nagaosa, and Y. Tokura, Nature **465**, 901 (2010).

7. A. Fert, V. Cros, and J. Sampaio, Nat. Nanotech. **8**, 152 (2013).

8. I. Dzyaloshinsky, J. Phys. Chem. Solids **4**, 241 (1958).

9. T. Moriya, Phys. Rev. **120**, 91 (1960).

10. X. Z. Yu, Y. Onose, N. Kanazawa, J. H. Park, J. H. Han, Y. Matsui, N. Nagaosa, and Y. Tokura, Nature **465**, 901 (2010).

11. A. Sonntag, J. Hermenau, S. Krause, and R. Wiesendanger, Phys. Rev. Lett. **113**, 077202 (2014).

12. S. D. Pollard, J. A. Garlow, J. Yu, Z. Wang, Y. Zhu, and H. Yang, Nat. Commun. **8**, 14761 (2017).

13. T. Schulz, R. Ritz, A. Bauer, M. Halder, M. Wagner, C. Franz, C. Pfleiderer, K. Everschor, M. Garst, and A. Rosch, Nat. Phys. **8**, 301 (2012).

14. J. Zang, M. Mostovoy, J. H. Han, and N. Nagaosa, Phys. Rev. Lett. **107**, 136804 (2011).

15. F. Jonietz, S. Muhlbauer, C. Pfleiderer, A. Neubauer, W. Münzer, A. Bauer, T. Adams, R. Georgii, P. Böni, R. A. Duine, K. Everschor, M. Garst, and A. Rosch, Science **330**, 1648 (2010).

16. Y. H. Liu, and Y. Q. Li, J. Phys.: Condens. Matter **25**, 076005 (2013).

17. M. Mochizuki, Phys. Rev. Lett. **108**, 017601 (2012)

18. Y. Onose, Y. Okamura, S. Seki, S. Ishiwata, and Y. Tokura, Phys. Rev. Lett. **109**, 037603 (2012).

19. S.-Z. Lin, C. D. Batista, and A. Saxena, Phys. Rev. B **89**, 024415 (2014).

20. A. Siemens, Y. Zhang, J. Hagemeister, E. Y. Vedmedenko and R. Wiesendanger, New J. Phys. **18**, 045021 (2016).

21. Y. Okamura, F. Kagawa, M. Mochizuki, M. Kubota, S. Seki, S. Ishiwata, M. Kawasaki, Y. Onose, and Y. Tokura, Nat. Commun. **4**, 2391 (2013).

22. D. Ehlers, I. Stasinopoulos, V. Tsurkan, H.-A. Krug von Nidda, T. Fehér, A. Leonov, I. Kézsmárki, D. Grundler, and A. Loidl, Phys. Rev. B **94**, 014406 (2016).

23. Y. H. Liu, Y. Q. Li, and J. H. Han, Phys. Rev. B **91**, 100402(R) (2013).

24. H. Yang, A. Thiaville, S. Rohart, A. Fert, and M. Chshiev, Phys. Rev. Lett. **115**, 267210 (2015).

25. K. Di, V. L. Zhang, H. S. Lim, S. C. Ng, M. H. Kuok, J. Yu, J. Yoon, X. Qiu, and H. Yang, Phys. Rev. Lett. **114**, 047201 (2015).

26. K. Di, V. L. Zhang, H. S. Lim, S. C. Ng, M. H. Kuok, X. Qiu, and H. Yang, Appl. Phys. Lett. **106**, 052403 (2015).

27. A. A. Stashkevich, M. Belmeguenai, Y. Roussigne, S. M. Cherif, M. Kostylev, M. Gabor, D. Lacour, C. Tiusan, and M. Hehn, Phys. Rev. B **91**, 214409 (2015).

28. O. Boulle, J. Vogel, H. Yang, S. Pizzini, C. D. de Souza Chaves, A. Locatelli, *et. al.*, Nat. Nanotech. **11**, 449 (2016).

29. M. Belmeguenai, J. P. Adam, Y. Roussigne, S. Eimer, T. Devolder, J. V. Kim, S. M. Cherif, A. Stashkevich, and A. Thiaville, Phys. Rev. B **91**, 180405(R) (2015).

30. J. Torrejon, J. Kim, J. Sinha, S. Mitani, M. Hayashi, M. Yamanouchi, and H. Ohno, Nat. Commun. **5**, 4655 (2014).

31. A. Vansteenkiste, J. Leliaert, M. Dvornik, M. Helsen, F. Garcia-Sanchez, and B. Van Waeyenberge, AIP Advances **4**, 107133 (2014).

32. H. Yang, O. Boulle, V. Cros, A. Fert, and M. Chshiev, "Controlling Dzyaloshinskii-Moriya interaction via chirality dependent layer stacking, insulator capping and electric field," arXiv:1603.01847v2 (2016).

33. Y. Zhou, E. Iacocca, A. A. Awad, R. K. Dumas, F. C. Zhang, H. B. Braun, and J. Akerman, Nat. Commun. **6**, 8193 (2015).

34. N. Romming, A. Kubetzka, C. Hanneken, K. von Bergmann, and R. Wiesendanger, Phys. Rev. Lett. **114**, 177203 (2015).

35. M. Mruczkiewicz, M. Krawczyk, and K. Y. Guslienko, Phys. Rev. B **95**, 094414 (2017).

36. M. Garst, J. Waizner, and D. Grundler, "Collective spin excitations of helices and magnetic skyrmions: review and perspectives of magnonics in non-centrosymmetric magnets," preprint arXiv:1702.03668v1 (2017).